\documentclass[a4paper,11pt]{article}
\usepackage{pos}
\usepackage{wrapfig}

\title{Using strong gravitational lensing to zoom in on high-redshift galaxies}

\author*[a]{Cristiana Spingola}

\affiliation[a]{INAF $-$ Istituto di Radioastronomia,\\
 Via Gobetti 101, I$-$40129, Bologna, Italy}

\emailAdd{spingola@ira.inaf.it}

\abstract{The centres of galaxies are powerful laboratories to test the current $\Lambda$CDM model for structure formation and evolution. While these sub-galactic scales can be directly investigated in the local Universe, it is observationally extremely difficult to access them at high redshift. The combination of strong gravitational lensing and VLBI observations allows us to access these scales to study both the baryonic and the dark matter distribution at the largest distances. For example, it becomes possible to unveil complex mass density distribution of lensing galaxies, faint cold molecular gas reservoirs, offset and binary AGN candidates at $z>1$.  Currently, these detailed studies are limited by the small number of known radio-loud lensed sources. Wide-field VLBI observations may provide a viable way to search for many more radio-loud systems and test strategies in preparation for the future surveys with the next generation of interferometers.

 }

\FullConference{%
  *** European VLBI Network Mini-Symposium and Users' Meeting (EVN2021) ***\\
  *** 12-14 July, 2021 ***\\
  *** Online ***
}

\begin{document}
\maketitle

\section{Introduction}
\subsection{Small scale issues of the $\Lambda$CDM cosmological model}

The $\Lambda$ Cold Dark Matter (CDM) cosmological framework predicts features that are remarkably close to what observed on large ($\sim$Mpc) scales, from the presence of filaments and voids to the clustering of galaxies and the angular dependence of the Cosmic Microwave Background \citep[e.g.,][]{Springel2005, Vogelsberger2020}. Yet, it presents several challenges at explaining some observables on galactic and sub-galactic scales \citep{Bullock2017}. According to this framework, the building of galaxies is driven by a hierarchical process of merging, which is dominated by the dark matter component. During the merging, the dense centers of small dark matter halos (M $\leq10^8$ M$_{\odot}$) are expected to survive this process and, therefore, larger dark matter halos should host many substructures. For example, a Milky Way-sized galaxy should host $\sim 10^3$ sub-halos, but only a few tens of satellites have been observed so far \citep{Klypin1999, Moore1999, Nierenberg2016, Drlica2020}. This conflict prompted revisions of the dark matter component of the standard model, as for instance a self-interacting "warm" dark matter. Warm dark matter models predict fewer low mass sub-halos to survive the merger process \citep[e.g.,][]{Vogelsberger2014, Lovell2020}, therefore they provide a solution to this discrepancy \citep[e.g.,][]{Dekker2021}.

From a baryonic matter perspective, the cold molecular gas is a key observable for understanding if galaxies form and evolve across the cosmic time following the $\Lambda$CDM model predictions, because it is the fuel for both star formation and active galactic nuclei (AGNs) \citep[e.g.,][]{Krumholz2012}. The bulk of star formation in the Universe happened between $z=1$ and $z=3$ \citep{Madau2014}, but the low surface brightness of the CO (1--0) makes it challenging to detect it at those redshifts \citep{Carilli2013,Decarli2019}. Also, quantifying the impact of AGNs on their host galaxy, including the surrounding cold gas, is one of the main unresolved questions in structure formation and evolution, both observationally and theoretically \citep{Schaye2015, Morganti2017}. At $\gamma$-rays AGNs can show the most violent and energetic explosions in the Universe, but to date it is not exactly clear what their effect on the host galaxy is and where these powerful events occur \citep{LeonTavares2011, Fuhrmann2014, Orienti2018, Orienti2020, Kramarenko2021}.

Another intriguing consequence of the hierarchical merger process at the base of $\Lambda$CDM model is the formation of close pairs of supermassive black holes (SMBHs, \citep{Begelman1980}). It is not yet known what happens when the SMBHs merge, but these mergers are the main target for the $\mu$Hz--nHz gravitational wave telescopes, such as pulsar timing arrays and the Laser Interferometer Space Antenna \citep{BurkeSpolaor2019}. Because of their small angular separation, the search for these binary systems is mostly limited to the low-z Universe \citep{Wang2009, Fu2012, Comerford2013, Rubinur2019}, making it challenging to compare the predictions from cosmological simulations with observations \citep[e.g.,][]{RosasGuevara2019}.

\subsection{Cosmic telescopes at the radio wavelengths}

Gravitational lensing is a powerful tool to overcome the limits of current instruments, because it allows us to study in detail distant galaxies that otherwise would not be even detected.
This effect consists of the deflection of the light from a distant background source by a foreground mass distribution (called "lens", for example a galaxy \citep{Treu2010}). As a result, multiple magnified images of the same high-$z$ galaxy may be observed. In the \textsl{strong} galaxy-galaxy lensing regime, the lensed images are usually separated by 1--2 arcsec \citep{Browne2003, Myers2003}, rarely separated by more than 5 arcseconds \citep{Phillips2001,McKean2005, McKean2011}, as lenses are mostly single early-type galaxies \citep[e.g.,][]{Treu2006}. Early work on the search for lensing systems separated by less than 300 mas has produced a null result \citep[][]{Augusto2001}.
Sub-arcsecond angular resolution is, therefore, a crucial requirement for finding lensing systems. However, higher angular resolution is necessary to spatially resolve the inner structure of the lensed images, which are typically compact, but they contain the magnified and amplified view of the background galaxies. 

At optical/near-infrared wavelengths dust extinction can dim the lensed images, hiding a large number of lensing systems. Also, stars in the host galaxy in front of the lensed images produce microlensing of a point-like background source \citep[e.g.,][]{Schmidt2010}, which corrupts the light curves that can be used to study the source variability and, ultimately, estimate $H_0$. Finally, the gravitational effect of the critical sub-halos of mass of M $<10^8$ M$_{\odot}$ is at milliarcsecond level (the order of magnitude of the corresponding Einstein radius). 

For all these reasons, the ideal observing band to search and study gravitational lensing systems for addressing the small scale issues of $\Lambda$CDM framework is the \textsl{radio}. First, at the radio wavelengths it is possible to achieve milliarcsecond (mas) and sub-mas angular resolution by using \textsl{interferometry}. This excellent angular resolution allows us to clearly spatially resolve and study in detail the emission from the lensed images. 
The radio emission is not affected by the optical emission due to stars and dust in the lensing galaxy. Moreover, radio-loud background sources are typically not point-like, but they extend at least for a few parsecs in size, hence they are not subjected to microlensing. Therefore, it becomes possible to measure any intrinsic variation in the background source with high accuracy.

\begin{figure*}[!htbp]
    \centering
    \includegraphics[width=0.9\textwidth]{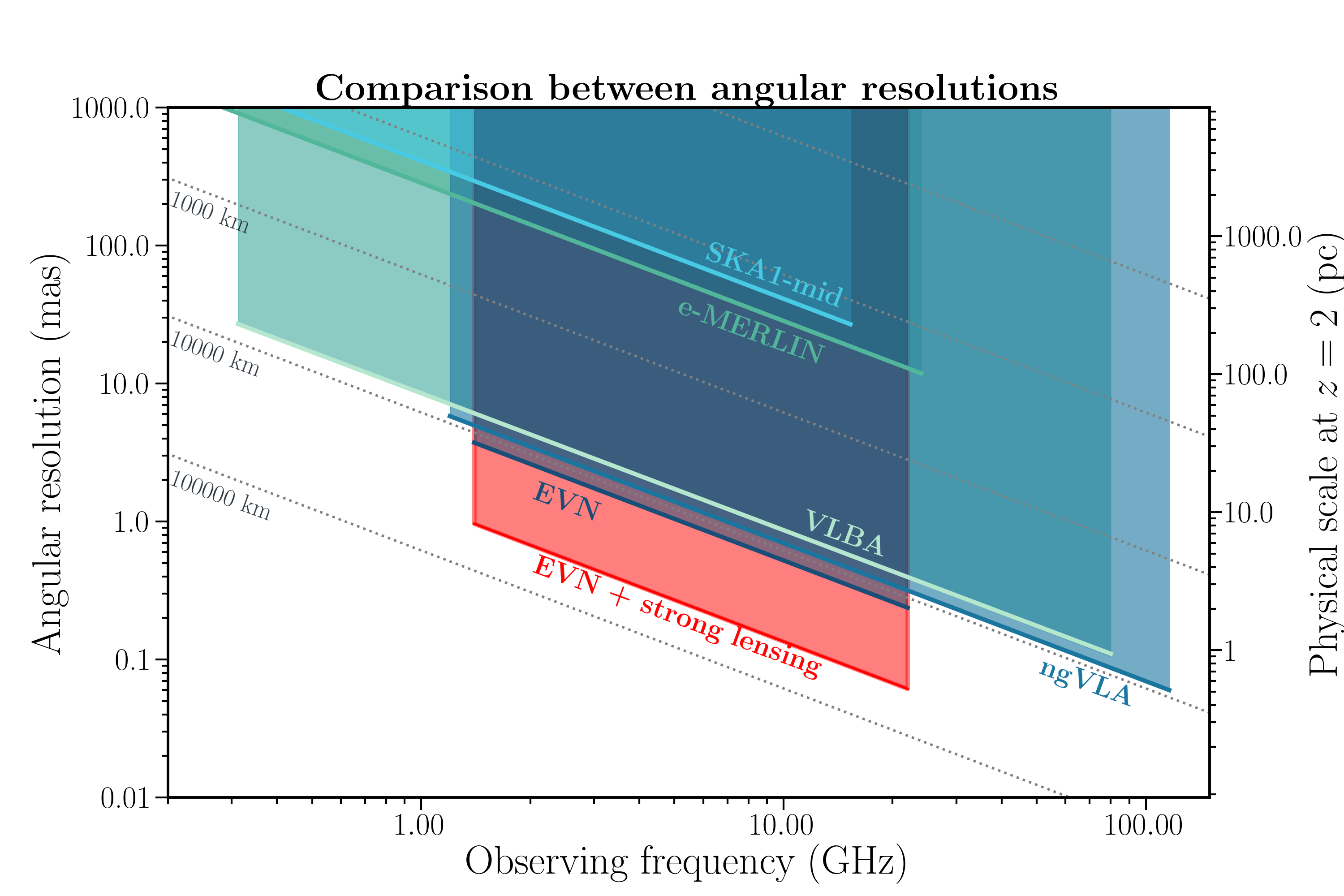}\\
    \includegraphics[width=0.9\textwidth]{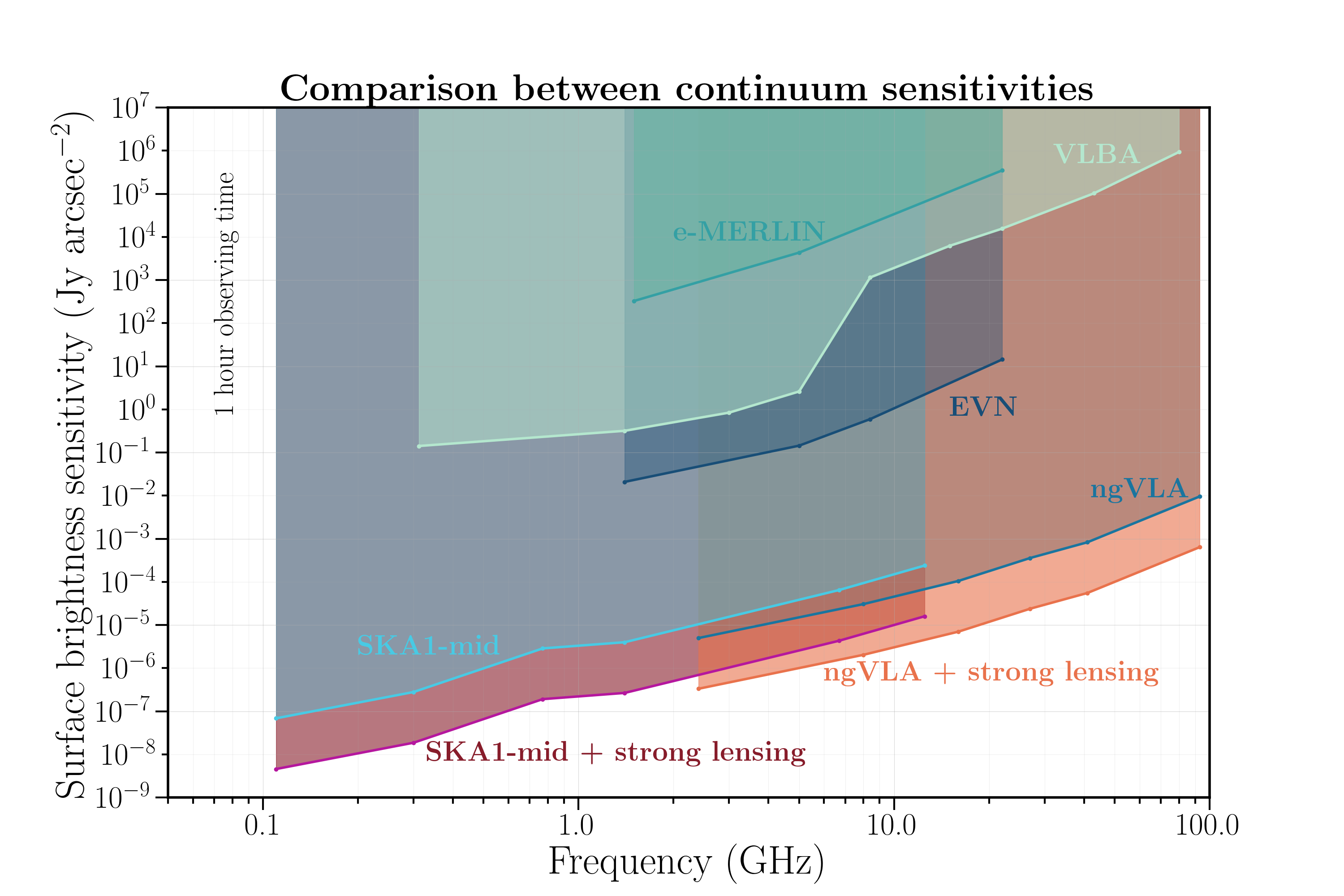}
    \caption{(Upper) Angular resolution as a function of frequency for some of the main radio interferometers that are operational (e-MERLIN, VLBA and EVN) or commissioned (SKA1-mid and ngVLA). The right y-axis shows the corresponding physical scale at $z=2$ ($8.568$ pc mas$^{-1}$ using the cosmological parameters of \citep{Planck2014}), which is epoch of the peak of the cosmic star formation and AGN activity. In red we show the gain in spatial resolution if the target source is gravitationally lensed by a magnification factor $\mu= 15$, which is a typical value for a quadruply imaged source. (Lower) Comparison between the point-like surface brightness sensitivity for 1 hour on-source for different arrays, assuming the largest bandwidth and best data recording rate for each array/band. In red and orange we show the boost in sensitivity that can be obtained by observing a gravitationally lensed source magnified by $\mu=15$ for the two most sensitive arrays at cm wavelengths (the SKA1-mid and the ngVLA).
    }
    \label{fig:lensing_vlbi_telescopes}
\end{figure*}

When radio interferometry is combined with strong lensing we can achieve the highest spatial resolution and sensitivity at the most distant cosmological epochs. In Fig. \ref{fig:lensing_vlbi_telescopes} we show how a typical magnification factor $\mu$ for a quadruply imaged source ($\mu = 15$) improves on angular resolution and sensitivity with respect to some radio interferometers (operational and commissioned). The sensitivity is improved by more than an order of magnitude for this case. This extremely high sensitivity enables the detection of the faintest (thus, more common) galaxies with a significantly reduced observing time. Also, observing a source that is magnified by $\mu = 15$ corresponds to adding a baseline of $50\,000$ km to the radio interferometer with the highest resolving power at cm wavelengths, which is the European VLBI Network (EVN) \citep{Venturi2020}. Therefore, the sources that are magnified the most are at a premium for detailed small-scale studies of dark and baryonic components in high-$z$ galaxies.

\section{Some applications of strong lensing and VLBI}

This Section provides an incomplete and radio-VLBI-biased overview of the applications of strong lensing. The most technical details are intentionally left out to give a more accessible review also for non-experts, but they can be found in the referenced papers. The order of the subsections follows a "top-down" approach, from the largest structures (kpc) to the smallest scales (pc).

\subsection{Dark matter distribution in lensing galaxies}

\begin{figure}
    \centering
    \includegraphics[width=0.9\textwidth]{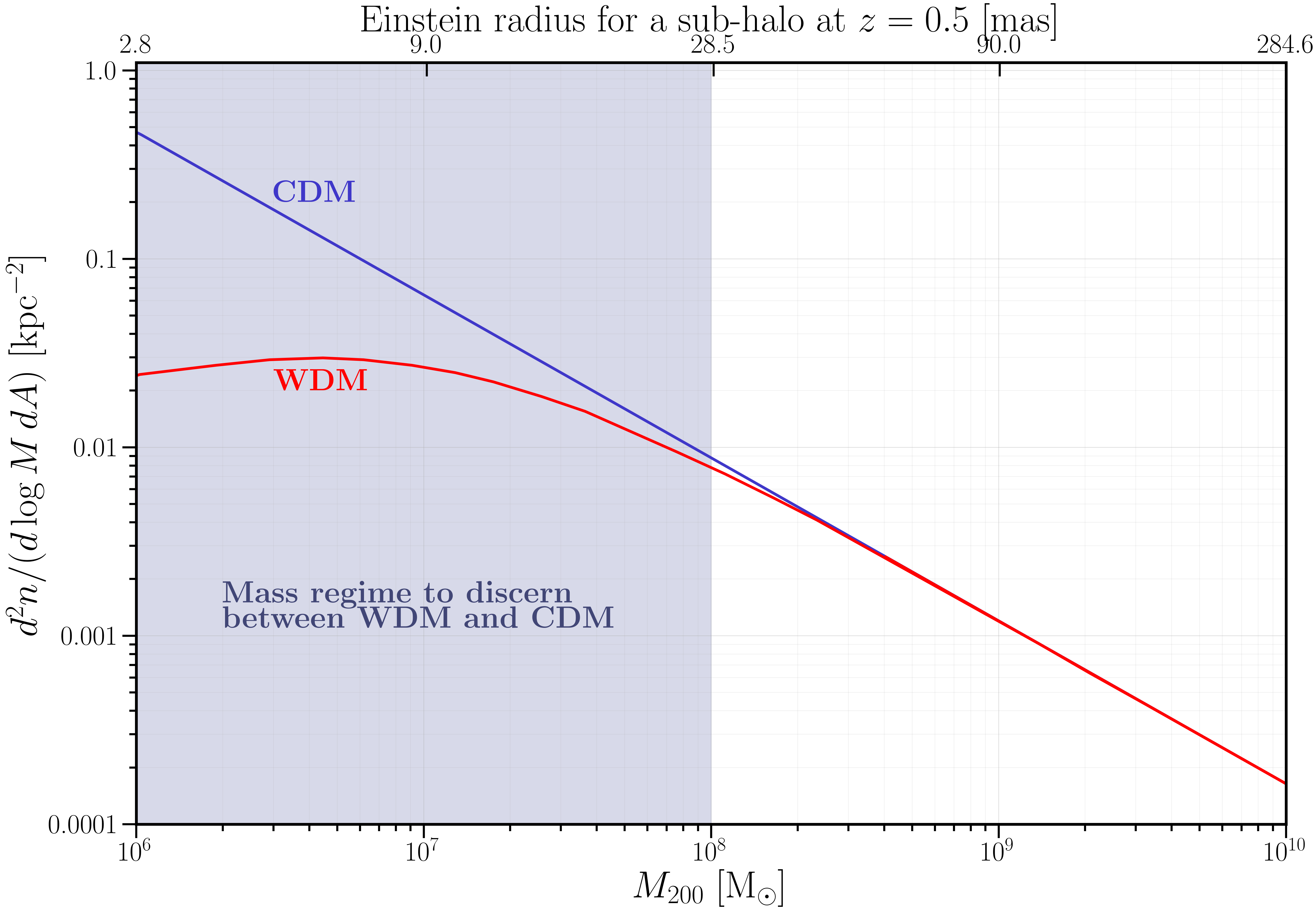}
    \caption{Sub-halo mass function for CDM and WDM models from \citep{Gilman2019}. Detecting sub-halos with mass between $10^6$ and $10^8$ M$_{\odot}$ would place strong contraints on the nature of dark matter \citep[e.g.,][]{Vegetti2012, Despali2018}. The gravitational effect of these low mass sub-halos would be observable on the scales of their Einstein radius, which we report on the top x-axis, assuming that they are singular isothermal spheres at $z=0.5$ (a typical redshift for lensing galaxies). VLBI arrays have the resolving power to observe the lensed images produced by such sub-halos (Fig. \ref{fig:lensing_vlbi_telescopes}).
    } \vspace{-0.7cm}
    \label{fig:cdm_subhalos_massfct}
\end{figure}

By examining anomalies (e.g., in their astrometry or flux/surface brightness) in the lensed images of strong gravitational lenses, it is possible to uniquely probe the presence of sub-halos at the same redshift of the lensing galaxy or along the line-of-sight \citep{Koopmans2005, Vegetti2012,  Despali2018, Gilman2019, Hsueh2020}. This is because the presence of a perturbation locally changes the magnification of one (or more) of the lensed images, consequently altering the observed flux density and position \citep{Metcalf2001}. The amplitude of the perturbation depends on several factors \citep{Despali2018}. The lensing systems that show many lensed images that are extended are ideal for these studies, as they provide a large number of constraints to the mass density distribution of the lensing galaxy and they maximize the probability to detect such perturbations \citep{Vegetti2012, Hezaveh2016, Ritondale2019, Spingola2018}. The $\mu$Jy beam$^{-1}$ surface brightness sensitivity of current VLBI arrays (Fig. \ref{fig:lensing_vlbi_telescopes}) allows us already to image at high fidelity the relatively extended emission of radio-loud lensed images while having mas angular resolution can theoretically enable the detection of possible perturbations due to the smallest sub-halos expected from the $\Lambda$CDM model, as shown in Fig. \ref{fig:cdm_subhalos_massfct}. 

With this aim, we started a campaign for observing a sample of strongly lensed AGN jets that can potentially show extended images and gravitational arcs.  As first source of this sample, we analysed and modelled the global VLBI observations (EVN and VLBA together) of the strong lensing system MG J0751+2716, which shows several extended arcs at an angular resolution of about 5 mas \citep{Spingola2018, Powell2021}. When parametrically modelling this rare lensing system with a smooth mass density distribution, we find offsets at 75$\sigma$ level on average between the observed and the model-predicted position of the lensed images, suggesting that there is either the need of extra mass (e.g., in the form of CDM sub-halos) or more complexity (e.g, angular structure) in the mass model. Further modelling of the visibilities with the novel method developed by \citep{Powell2021} is on-going. Grid-based potential corrections are needed to clearly assess the nature of these astrometric anomalies \citep{Koopmans2005, Vegetti2009}.

\subsection{Baryonic matter in high-$z$ lensed galaxies}

The extended lensed images contain the super-resolved and amplified information about the background galaxy. This is particularly important for studying the faintest emission from the cold molecular gas in high-$z$ galaxies with high detail. Thanks to the lensing magnification, we could spatially resolve the cold molecular gas and stellar content with sub-kpc angular resolution and high sensitivity in two strongly lensed jetted AGNs MG J0751+2716 and JVAS B1938+666 at redshift $z=3.2$ and  $z=2.0$, respectively. Both these systems appear as extended gravitational arcs both at optical wavelengths and radio continuum/CO (1--0) frequencies \citep[][]{Spingola2020_gas}. We reconstructed the optical source properties using the sophisticated grid-based lens modelling software developed by \citep[][]{Vegetti2009,Ritondale2019_code}. 
We modelled the radio continuum and CO (1--0) data directly in the visibility plane using the expansion of this software developed by \citep[][]{Powell2021}\footnote{We highlight that this technique for radio data has been pioneerly explored by \citep[][]{Wucknitz2004a, Wucknitz2004b}. }. As the software uses the entire surface brightness distribution of the arcs, we could infer the lens model parameters at high precision and reconstruct the multi-wavelength emission of the sources at high image fidelity. 
We find intrinsic velocity fields on kpc-scales showing distinct structures that can be disks (elongated velocity gradients) and possibly interacting objects (off-axis velocity components). In both sources there is an offset between the peak of the CO (1--0) intensity and the peak of the radio continuum emission, which could be an indication of radiative feedback from the AGN (or offset star formation). Also, these two galaxies show at their centre very compact stellar cores ($<$1 kpc). Altogether, these lensed source provide a picture that is well in agreement with the current structure formation model, which predicts gas fractions larger than 30 \% and compact central stellar cusps to be already in place at redshifts $z\sim1-2$ \citep[e.g.,][]{Conselice2007}. Therefore, the detailed study of gas and stars in lensed galaxies can be a valid way to test specific evolutionary stages predicted by hydro-dynamical simulations (see also \citep[][]{Stacey2021}).

\subsection{Searching for gravitationally lensed binary and offset AGNs}

The dominant formation mechanism of AGN is still under debate (i.e., merging of multiple SMBHs versus accretion of the surrounding material). Therefore, assessing the rate at which binary AGN systems occur provides strong constraints on the role of merging in the AGN formation and evolution. By comparing two VLBI observations separated by $\sim 14$ years
of the radio-loud gravitationally lensed AGN MG B2016+112 ($z_{\rm source}$ = 3.273) we measured for the first time proper motions in the complex source plane \citep{Spingola2019_prop_motion}. The proper motion and multi-band properties of this system suggest that the source might consist of two radio-loud AGNs separated by only $\sim175$ pc in projection \citep{Spingola2019_prop_motion}. If confirmed, this discovery would suggest that such binary AGN systems must be more abundant in the early Universe than currently estimated \citep[e.g.,][]{RosasGuevara2019}. In order to better characterize this system and understand its nature, we expanded this analysis to the X-ray band using \textsl{Chandra} observations \citep{Schwartz2021}. We find that the X-ray emission from this AGN is consistent with two X-ray sources, adding more evidence in favour of the binary AGN scenario.

\begin{figure}
    \centering
    \includegraphics[width=0.9\textwidth]{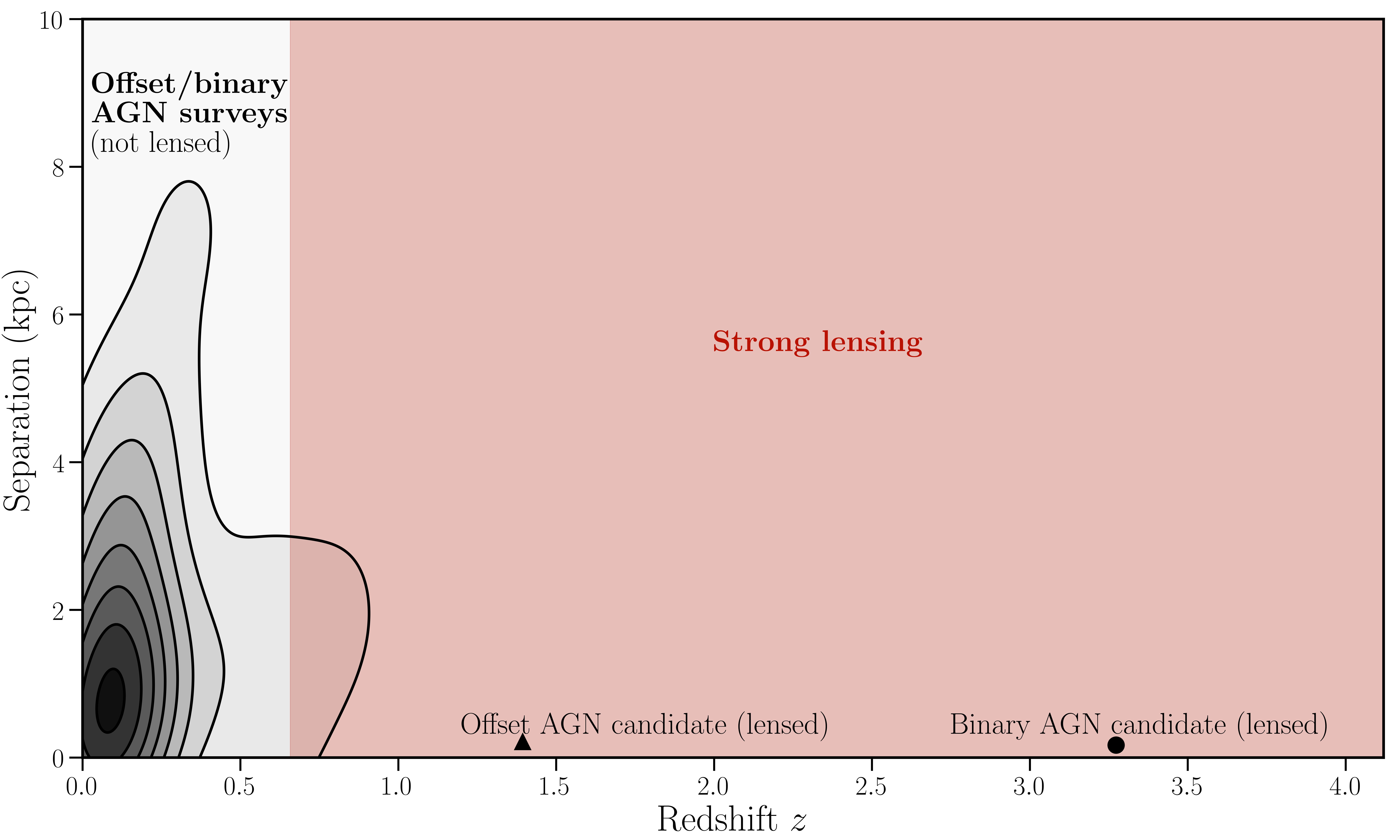}
    \caption{Average separation (kpc) as a function of redshift for the largest binary and offset AGN searches (adapted from \citep{Spingola2021_submitted}). With \textsl{separation} we refer to the distance between the optical peak of the emission of the host galaxy and radio/X-ray emissions (for offset AGNs) or the angular separation between the two AGNs (for binary AGNs). We plot using a black triangle and a black circle the lensed offset AGN candidate CLASS~B1608+656 \citep{Spingola2020} and the lensed binary AGN candidate MG~B2016+112 \citep{Spingola2019_prop_motion, Schwartz2021}, respectively. The lensing magnification allows us to spatially resolve the smallest systems at the highest redshifts. The red shaded area indicate the redshift range covered by lensing systems with precise lens mass models, radio, optical and X-ray \textsl{Chandra} observations.}
    \label{fig:offset_binary_surveys}
\end{figure}

Motivated by these findings, we started the first systematic search for binary AGN systems using a strong lensing sample of systems that have radio VLBI, optical Hubble Space Telescope (HST) and X-ray \textsl{Chandra} observations. The most promising lensed sources for finding close binary AGNs are those that are quadruply imaged, because of their high magnification. Also, the lens mass model provides a universal reference frame, allowing us to robustly measure the relative position of the multi-wavelength emission \citep{Barnacka2017, Barnacka2018}. We selected a pilot sample of lensing systems with precise VLBI-derived mass models and, among them, we found an offset AGN candidate at $z=1.394$ with an optical-radio offset of $25\pm16$ mas ($214\pm137$ pc in the source frame \citep{Spingola2020}). An offset AGN could be a binary AGN system where only one SMBH is active. Therefore, its radio (or X-ray) emission appears offset with respect to the peak of the optical emission of its host galaxy \citep[e.g.,][]{Comerford2014}. However, the nature of offset AGNs is complex, as the displacement of a SMBH could be also produced by the merger process that "kicks" the SMBH in a position that is not at the center of the galaxy \citep[][]{2014Lena}. Therefore, also a population of offset SMBHs is expected: the combination of observations of lensed binary and offset AGNs has the potential strongly constrain the current SMBHs formation models \citep[][]{Volonter2003, Ricarte2021}. 

\subsection{Locating the most powerful flares at $\gamma$-ray energies}

Time delays are among the primary manifestations of gravitational lensing. 
Light rays coming from the background source travel paths with different lengths towards the observers: as a result, there is a time difference in the arrival time of the signal (known as time delay $\Delta t$) at each lensed image. Time delay measurements are typically used to infer the Hubble parameter $H_0$ \citep[e.g.,][]{Barnacka2015_h0, Grillo2018, Chen2019}, as $\Delta t$ is inversely proportional to $H_0$ \citep{Refsdal1964}. At the radio wavelengths the monitoring of each lensed image is easy to perform (but time expensive), and it provides robust flux densities to determine gravitational time delays (also using variations in polarization \citep[e.g.,][]{Biggs2021}). Nevertheless, for such estimate there is the need of strong features in the light curves \citep[e.g.,][]{Rumbaugh2015}.

Comparing the gravitationally-induced time delays at different wavelengths is also an effective method to spatially locate the emitting regions in high-$z$ sources. For instance, a difference between  the measured radio and the $\gamma$-ray time delays directly implies that the two regions are at a different location in the source plane. This is particularly powerful at $\gamma$-ray energies, where the poor angular resolution does not allow to spatially resolve the lensed images and, therefore, it precludes any precise localization of the high energy emitting region at high-$z$ \citep[e.g.,][]{Barnacka2016}. However, there are only two known lensed AGNs that are $\gamma$-ray emitters, JVAS B0218+357 and PKS 1830-211 \citep{Cheung2014, Barnacka2015, Spingola2016}.

\begin{wrapfigure}{l}{0.6\textwidth}
    \begin{center}
        \includegraphics[width=0.6\textwidth]{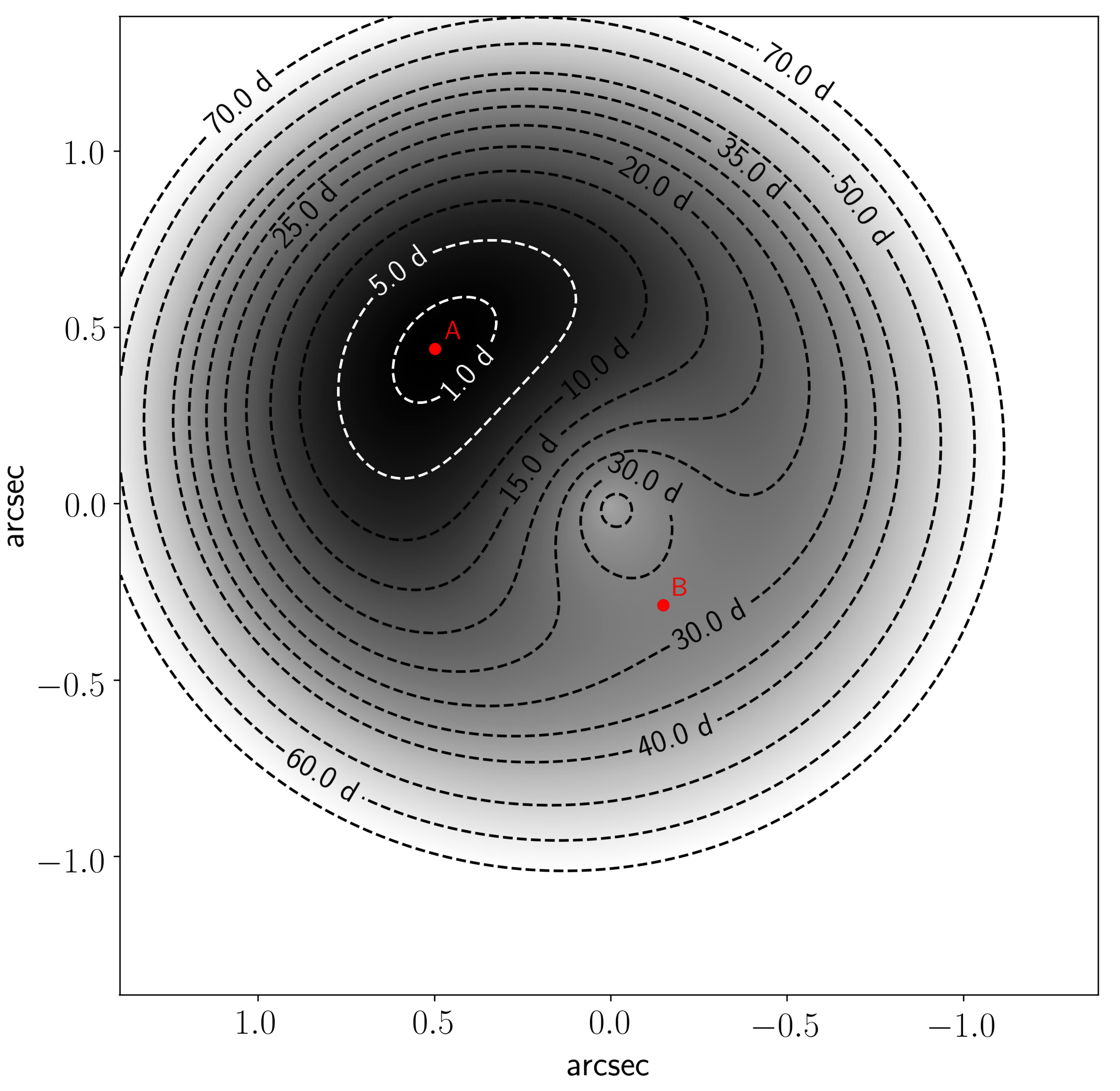}
    \vspace{-0.5cm}
    \caption{Fermat time delay surfaces (grey scale and dashed lines, in days) for the doubly imaged blazar PKS 1830-211 \citep{Spingola2022}. The red circles indicate the position of the lensed images A and B. Image A has "zero" time delay because the signal arrives there before appearing in the lensed image B (where the signal is delayed).}\label{fig:time_delay}
        \end{center} \vspace{-0.5cm}
\end{wrapfigure}  

With the aim of better constraining the radio/$\gamma$-ray connection in flaring AGNs, we monitored for about $\sim3$ months the doubly imaged blazar PKS 1830-211 at the radio wavelengths.  In April 2019 this lensed source underwent its brightest and longest outburst at $\gamma$-rays, and a very high state of activity was detected also at the radio wavelengths with the EVN antenna of Medicina \citep{Atel2019a, Atel2019b}. In order to estimate the radio time delay, we followed-up this system at mas angular resolution with multi-frequency observations with the VLBA right after the $\gamma$-ray flare. We obtained spatially resolved light curves at 15, 22 and 43 GHz. The best (lowest $\chi^2$) time delay value for the 43 GHz light curve is $\Delta t = 28.15 \pm 0.76$, which is a significant improvement on the precision with respect to the previous radio time delay measurements for this system \citep[][]{Lovell1998} and in well agreement with the lens mass modelling predictions of \citep[][]{Muller2020}. Using this estimate of the time delay and the VLBI positions of the lensed images, we refined the lens mass model, which is shown in Fig. \ref{fig:time_delay} as time delay surfaces. Our model can successfully reproduce the VLBI and the ALMA lensed images positions from \citep[][]{Muller2020}. We look forward to comparing our measurement at the radio wavelengths with the time delay at $\gamma$-rays. For instance, a difference of 2 days in time delays corresponds to about 160 parsecs distance in projection between the radio and the $\gamma$-ray emitting regions at the redshift of the source $z=2.507$.

\section{An outlook to the future}

The study of radio-loud gravitationally lensed sources at mas resolution seems a promising way to address the small scale issues of $\Lambda$CDM, but it comes at a price: to date, we know only $\sim35$ radio-loud lensing systems. Despite this small number, such objects provide a wealth of information on the smallest angular scales of both the foreground lenses and the background sources, which would not be possible to access at high redshift otherwise. But, to put statistically significant constraints on the small scales of sources and lenses, it becomes clear that there is the need to find many more radio-loud gravitational lensing systems, possibly both with compact and extended lensed images. To do so, we used for the first time the wide-field VLBI survey mJIVE-20 \citep{Deller2014}. Using multi-band observations and simple selection criteria based on lensing properties, we found two known radio-loud lensing systems and two more lens candidates, which we followed up with sensitive EVN+eMERLIN observations to confirm their nature \citep[][]{Spingola2019_mjive}. Another recent lens search using VLBI data from the Astrogeo survey has been presented at this conference (see \citep[][]{Casadio2021}). In addition to the typical lensing criteria, they also used a citizen science approach and they are following-up the lens candidates with the EVN. Therefore, searching lenses in wide-field VLBI surveys provides a feasible way to find new systems already, also thanks to the significant advance in calibration and self-calibration of such data, which can allow one to detect the extremely faint (sub-$\mu$Jy level) radio population \citep[e.g.,][]{Radcliffe2016, Hartley2019, Badole2020}.

In the foreseeable future the wide-field all-sky surveys with the next generation of radio facilities, such as the Square Kilometer Array (SKA), are predicted to find $\sim 10^5$ lensing systems \citep{McKean2015}. If the excellent SKA sensitivity will be complemented by a VLBI capability \citep[][]{Paragi2015}, this large sample will provide an ideal statistical (and unbiased) sample to fully characterize the small scales of distant galaxies.

\section*{Acknowledgments}
CS acknowledges financial support from the Italian Ministry of University and Research - Project Proposal CIR01\_00010. 

\bibliographystyle{JHEP}
\bibliography{refs}

\end{document}